\documentclass{eptcs}
\usepackage{graphicx}
\usepackage[T1]{fontenc}   
\usepackage{amsmath,amsfonts,amssymb,mathrsfs}
\usepackage{relsize}
\usepackage{latexsym}
\usepackage{txfonts}
\usepackage{mathptmx}
\usepackage{breakurl}

\providecommand{\event}{F-IDE 2016} 

\makeatletter
\newcommand{\@chapapp}{\relax}%
\makeatother

\usepackage{microtype}            

\newcommand{\cxsemfreccia}[1]{{\overset{#1}{\longrightarrow}}}

\def\point(#1){({\it #1\/})}

\title{Extending a User Interface Prototyping Tool with Automatic
MISRA~C Code Generation}

\author{Gioacchino~Mauro
\institute{Department of Information Engineering,
University of Pisa, Pisa, Italy}
\email{giocchi27@gmail.com}
\and Harold~Thimbleby
\institute{Swansea University --- Prifysgol Abertawe, Swansea/Abertawe, UK}
\email{harold@thimbleby.net}%
\and Andrea~Domenici \qquad\qquad Cinzia~Bernardeschi
\institute{Department of Information Engineering,
University of Pisa, Pisa, Italy}
\email{\{cinzia.bernardeschi,andrea.domenici\}@unipi.it}
}

\def\titlerunning{MISRA~C code generation}
\def\authorrunning{G. Mauro, H. Thimbleby, A. Domenici \& C. Bernardschi}

\begin{document}

\maketitle

\begin{abstract} 
We are concerned with systems, particularly safety-critical systems, that
involve interaction between users and devices, such as the user interface of
medical devices. We therefore developed a MISRA~C code generator for formal
models expressed in the PVSio-web prototyping toolkit.
PVSio-web allows developers to rapidly generate realistic interactive
prototypes for verifying usability and safety requirements in human-machine
interfaces.  The visual appearance of the prototypes is based on a picture
of a physical device, and the behaviour of the prototype is defined by an
executable formal model.
Our approach transforms the PVSio-web prototyping tool into a model-based
engineering toolkit that, starting from a formally verified user interface
design model, will produce MISRA~C code that can be compiled and linked
into a final product.
An initial validation of our tool is presented for the data entry system of
an actual medical device.

\end{abstract}

\begin{figure*}[t]
\centering
\includegraphics[width =0.9\linewidth]{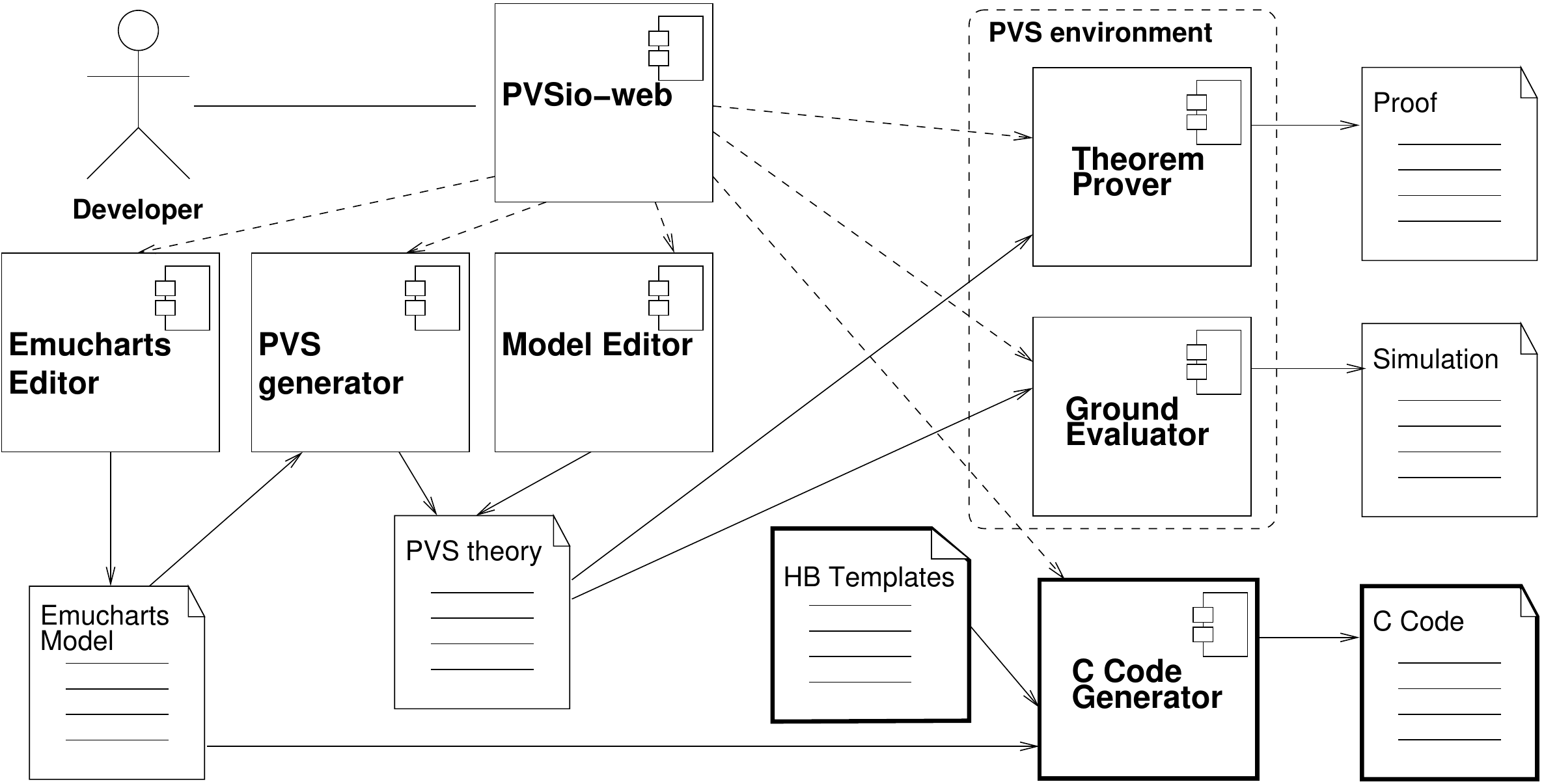}
\caption{C code generation in the PVSio-web development process.}
\label{fig:diagram}
\end{figure*}

\section{Introduction}

Formal methods are important for developing and
understanding safe and secure systems.  The PVSio-web
framework~\cite{masci-verisure2015,masciFHIES:14,masci-CAV2015,%
fmis2013-pvsioweb}
allows developers to use formal methods in a friendly and
appealing way as it provides realistic animations and is integrated with a
graphical editor for the Emucharts language \cite{masci:NFM2014}. (Emucharts is
a state machine formalism with guards and actions associated with transitions;
it is explained further in Sect.~\ref{Emucharts} below.)

PVSio-web uses the formal modelling language of the Prototype Verification
System (PVS)~\cite{Owre:1992}, including the PVSio extension~\cite{munoz03}.
PVS is an industrial-strength theorem proving system that allows formal
verification of safety and reliability properties of hardware and software
systems~\cite{Bernardeschi2008,srivas97}.  Although PVS itself is very
effective, it is not widely used for model-based development and analysis of
user interfaces, as the tool has a steep learning curve.  PVSio-web softens
this learning curve, making the tool more user-friendly and accessible,
providing developers with a graphical modelling environment, and a toolbox
for developing realistic visual prototypes of user interfaces.

The applications of embedded software in safety-critical applications increase
continuously.  Taking this into account, together with the requirements to
reduce time and overall production costs, automatic code generation plays
an essential role.  Automatic code generation guarantees a smooth conversion
from model to code and reduces the debugging and testing required for source
code, provided that the correctness of code generation and of the high-level
model have been verified.

We therefore present an extension to PVSio-web that generates C code.
Specifically, our extension generates MISRA~C~\cite{misra1}, a safety-oriented
subset of C developed by the Motor Industry Software Reliability Association
(MISRA)\@.  MISRA~C is commonly used in safety-critical subsystems, such as
car braking in automotive systems.

With this new extension, formal PVS specifications generated from Emucharts
diagrams are automatically converted to C, significantly shortening project
development time.  Because of the approach, the semantics of generated C
code is equivalent to the formal models, and therefore the code retains the
reliability and safety properties formally verified for the PVS model.

In summary, our main contribution is an approach to software development that
integrates logic- and state machine-based formal modelling, validation by
simulation, and automatic implementation by generating production code to
be run on the actual system hardware, all based on an industrial-strength
formal methods toolkit.

\section{Related work}
\label{rw}

Model-based approaches are commonly used in the field of human computer
interaction, for example \cite{Foley94}.  Most approaches are focused on
describing user interfaces and their implementations at various levels
of abstraction.  Developers of user interfaces for interactive systems
also have to address heterogeneity and adaptation to the context of use.
For example, in \cite{Paterno09}, a model-based declarative language for
the design of interactive applications based on Web services in ubiquitous
environments was presented.  In contrast to these familiar approaches, the
present work proposes a framework enabling a formal verification of user
interaction.  The framework is meant for safety analysis of safety-critical
devices and not with user interface design issues as discussed
in~\cite{Paterno09},

Similarly to our approach, formal models were used in \cite{Bowen11}
to describe functionality and component interactions, where they were
combined with user interface models in order to get the entire model of the
system. Moreover, an Android emulator application was generated, using Java
and XML technologies.  Presentation models and presentation interaction
models were used in \cite{Bowen07} to model interactive software systems;
these models were shown to be usable with a formal specification of the
system functionality.  In \cite{Bowen12} the same formalisms  were used to
model user manuals of modal medical devices, proving that the user manual
may be not always consistent with actual device behaviour.

In \cite{Harrison01}, model checking was used to model and prove
properties of specifications of interactive systems so that possibly unexpected
consequences of interface mode changes can be checked early in the design
process.  In \cite{Harrison13}, the complementary role of model checking and
theorem proving in the analysis of interactive devices was considered.
Recent work \cite{Harrison16} explored the paths that a user will take
in interacting with medical devices for the analysis of properties of the
behaviour of safety-critical  devices.  A model-checking approach has also been
used to analyse hardware behaviour~\cite{GLSVLSI}.

\begin{sloppypar}
A discussion of production code generation in model-based development can be
found in~\cite{erkkinen07}.  Many papers deal with specific code generators,
for example TargetLink~\cite{beine04}.  Code generators specifically designed
for medical systems are described in~\cite{banerjee14} and~\cite{pajic14}.
\end{sloppypar}

\section{PVS, model-driven development and Emucharts}
\label{bkgnd}

This section provides background information on the PVSio-web framework and
its relationship to model-driven development.

\subsection{PVS, the Prototype Verification System}

The PVS is an interactive theorem prover for a typed higher-order logic
language, providing an extensive set of inference rules based on the sequent
calculus~\cite{smullyan95}.  Its PVSio extension is a ground evaluator
that can compute the results of ground function applications, that is PVS
expressions consisting of a function name applied to variable-free arguments.
PVS functions are purely declarative definitions of mathematical mappings,
without any procedural information on how to compute them, but the PVSio
package can derive and execute an algorithm to evaluate a ground function
application, turning it into a procedure call.  The PVSio package also
provides functions with side effects, such as input and output, which do
not interfere with the semantics of a theory.

A system is modelled in PVS as a \emph{theory}, a collection of logical
statements and definitions about the structural and behavioural aspects of the
system.  The system's required properties are expressed as theorems to be
verified with the PVS theorem prover.  If the behavioural aspects are
expressed as functions, the system can also be simulated with the PVSio
extension.  The same logical model can then be used both for verification and
simulation.

\subsection{Model-driven development}

Model-driven development (MDD) is based on creating an executable system model
by assembling functional blocks.  An executable model makes it possible both
to simulate the system and to generate production software to control it.
Together with the naturalness of the graphic language of functional blocks,
these features make MDD very attractive to developers.  However, this approach
has two limits: first, functional blocks lend themselves to building design
models, but not specification ones; and secondly, formal verification of
a block-based model is tedious, and in fact it is uncommon in industrial
practice.

A formal approach can be used to create both specification and design models
and intrinsically lends itself to rigorous verification of system properties.
In particular, logic specification languages, such as PVS, are supported
by automatic or interactive theorem provers used by developers to check
if system requirements, expressed as logical formulas, are implied by a
system's description expressed in a logic theory.  However, formal methods
require expertise in languages and methods that are not widely known in the
wider developer community.  Further, most formal languages abstract from
the familiar procedure-oriented computation model of popular programming
languages, making it harder to generate executable software.

It is then desirable to have tools and methods providing developers with the
features of both approaches.  The present work is part of a research effort
aimed at this goal.  With the PVSio-web framework, a developer can build a
model in a graphical state-machine language or a logic language, or both
(Sect.~\ref{bkgnd}).  The graphical model is translated into the logic
language automatically, and the resulting translation is both verifiable and
executable using the PVSio ground evaluator, which acts as an interpreter
for the PVS language.  The PVSio-web framework thus provides features of the
formal approach: A formal specification language and a verification tool,
and features of MDD, thus providing a full graphical modelling language and
a simulation engine.  A translator from Emucharts to C makes it possible
to generate code from a state machine-based model that can be validated
by simulation and verified by theorem proving. The other important feature
of MDD --- generation of production code capable to be run on the actual
system hardware --- is a key contribution of this paper.

\subsection{PVSio-web}

The PVSio-web framework is a set of tools, co-ordinated by a web-based
interface, for prototyping and simulation of interactive devices.  Its main
components are, besides PVS with its PVSio
extension: 
\point(i) the \textbf{Prototype Builder}, a graphical tool used to choose
a picture of an existing or anticipated device's front panel and to associate
PVS functions with active areas of the picture representing device inputs
(e.g., buttons or keys) and outputs (e.g., alphanumeric displays or lights);
\point(ii) the \textbf{Model Editor}, a textual interface to write PVS code; 
\point(iii) the \textbf{Emucharts Editor}, a graphical tool to draw Emucharts
state machine diagrams; 
\point(iv) a \textbf{Simulation Environment}; and 
\point(v) \textbf{Code Generators} for PVS and other formal languages
(currently Presentation Interaction Models~\cite{bowen2015design},
Modal Action Logic~\cite{harrison2015reusing}, and Vienna Development
Method~\cite{fitzgerald05}) --- and for MISRA~C, as presented in this paper.

PVSio-web can be used to prototype a new device interface, or to create a
reverse-engineered model of an existing one.  In either case, a developer
creates formal descriptions of the device's responses to user actions, using
the model and Emucharts editors, and associates these descriptions with the
active areas of the simulated interface, using the prototype builder.  In the
simulation environment, the developer, or a domain expert or a potential user,
interacts with the prototype clicking on the input widgets.  These actions
are translated to PVS function calls executed by the PVSio interpreter.

\subsection{Emucharts}\label{Emucharts}

An Emucharts diagram is the representation of an extended state machine
in the form of a directed graph composed of labelled \emph{nodes\/}
and \emph{transitions}.  Transitions are labelled with triples of the
form $\mathit{trigger} [ \mathit{guard} ] \{ \mathit{action} \}$, where
\emph{trigger\/} is the name of an event, \emph{guard} is an enabling Boolean
expression, and \emph{action} is a set of assignments to typed {variables}
declared in the state machine's \emph{context}.  The default guard is the
\emph{true} value and the default action is a no-operation.  The \emph{state}
of the machine is defined by the current node and the current values of the
context variables.

The code generator for PVS produces a theory containing functions that define
the state machine behaviour on the occurrence of trigger events.  Since an
Emucharts diagram usually represents a device response to user actions, such
events represent user actions, such as pressing a button on a control panel.
During simulation on a PC, a user click on an active area of the device
picture causes the simulator to generate a function application expression
that is passed to the PVSio ground evaluator.

\section{From Emucharts to safe C}
\label{emupvs}

The aim of programming code generation in the PVSio-web framework is
producing a module that implements the user interface of a device, which
can be compiled and linked into the device software without any particular
assumptions on its architecture.  In this way, the user interface module
can be used without forcing design choices on the rest of the software.
In our approach, the generated module contains a set of C functions.
The main ones are, for each Emucharts trigger:
\point(i) a \emph{permission} function, to check if the
trigger event is \emph{permitted}, i.e., whether it is associated with any
transition from the current state, and 
\point(ii) a \emph{transition} function that, according to the current state,
updates it, provided that the guard condition of an outgoing transition holds.
The code includes logically redundant tests (\emph{assert\/} macros) to
improve robustness.

To generate production-quality code fit for safety-critical applications we
adopt MISRA guidelines.  The MISRA guidelines for the C language, originally
conceived for the automotive industry, enforce programming practices to
improve maintainability and portability and, above all, to reduce the risk
of malfunction due to implementation- or platform-dependent aspects of the
C language.  For instance, there are rules that bar the use of  constructs
such as \texttt{goto}, and rules requiring that numeric literals be suffixed
to indicate their type explicitly.  The generated code currently complies with
the first version of the 1998 MISRA~C guidelines.

\begin{figure*}[t]
\centering
\includegraphics[width=0.9\linewidth]{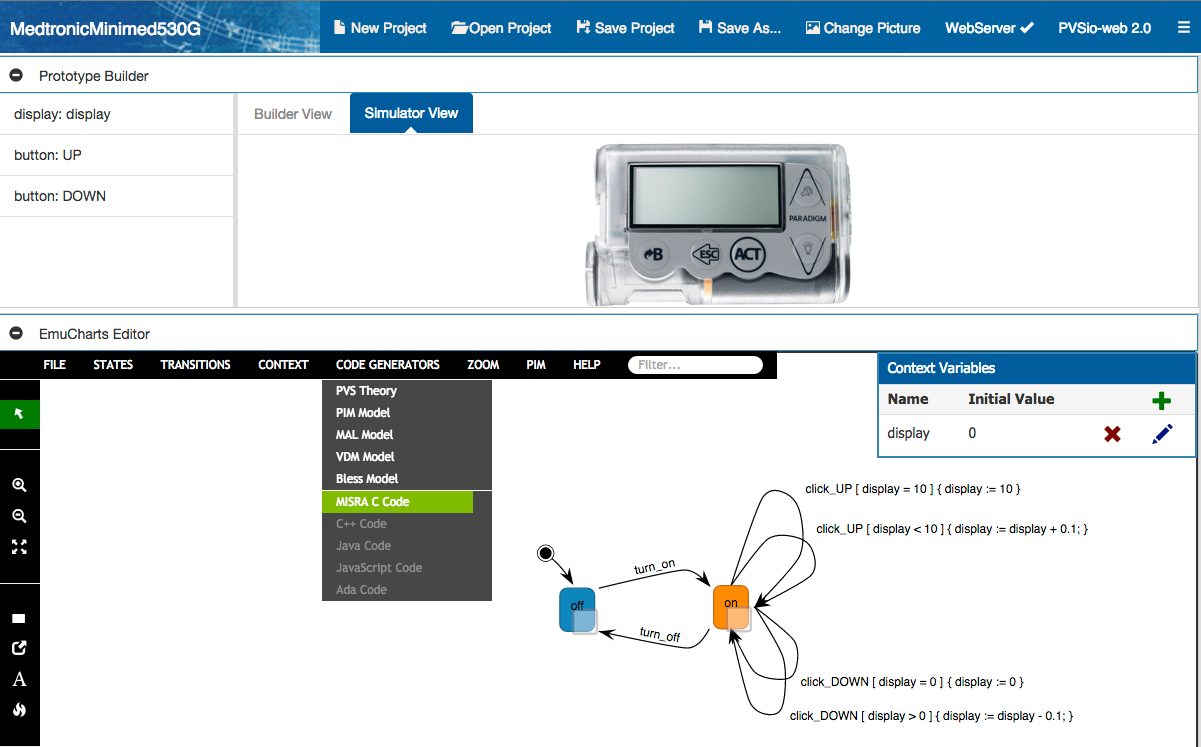}
\caption{The PVSio-web user interface with the Prototype Builder
and Emucharts Editor frames.}
\label{fig:Emucharts}
\end{figure*}

\subsection{Code generation}
\label{cgen}

\begin{sloppypar}
Our MISRA~C code generator was  implemented in JavaScript using
Handlebars~\cite{handlebars}, a macro-expansion tool for web applications.  A
Handlebars template is a piece of text containing ``{Handlebar expressions},''
which refer to elements of the surrounding context, typically an HTML document.
A Handlebars expression specifies a character string as a function of context
elements, which is compiled into a JavaScript function that returns the
template text with the substitutions computed by the Handlebars expressions.
For example, a template fragment for a C preprocessor \texttt{\#include}
directive is \texttt{\#include~"\{\{filename\}\}.h"}, where the Handlebars
expression \texttt{\{\{filename\}\}} contains the
\texttt{filename} parameter
that will be replaced by the actual name of the file to be included.
\end{sloppypar}

The code generator produces a header file, an implementation file, a makefile,
a simple test driver file, and a documentation manual.

The structure of the header file is defined by the grammar in
Table~\ref{initG}.  The header file contains, among other items, the
declarations (\emph{typedef\_definitions} in the grammar) for types
with explicit representation of size and sign, e.g., \texttt{UC\_8}, for
\emph{eight-bit unsigned char}, the declaration (\emph{state\_labels\_enum})
for an enumeration type defining the node labels, and the declaration
(\emph{state\_structure}) for the \emph{state} structure type representing
the state of the Emucharts model.  This structure contains one \emph{context}
field for each variable defined in the Emucharts context, and two more fields
(\emph{curr\_node} and \emph{prev\_node}) contain the labels of the current
and the previous node.

\begin{table}
 \begin{tabular}{r l l }
$\langle$ \emph{headerfile} $\rangle$ & ::=
          & $\langle$ \emph{preprocessor\_directives} $\rangle$ \\
        & & [ $\langle$ \emph{constant\_definitions} $\rangle$ ]\\
        & & $\langle$ \emph{typedef\_definitions} $\rangle$\\
        & & $\langle$ \emph{state\_labels\_enum} $\rangle$\\
        & & $\langle$ \emph{state\_structure} $\rangle$\\
        & & $\langle$ \emph{utility\_functions} $\rangle$\\
        & & $\langle$ \emph{init\_function} $\rangle$\\
        & & $\langle$ \emph{permission\_functions} $\rangle$\\
        & & $\langle$ \emph{transition\_functions} $\rangle$\\
        & &
 \end{tabular}
 \caption{Structure of a header file.  Non-terminal symbols are enclosed
  between angle brackets; square brackets enclose optional symbols.}
\label{initG}
\end{table}

\begin{figure}[tbh]
\centering
\includegraphics[width=\linewidth]{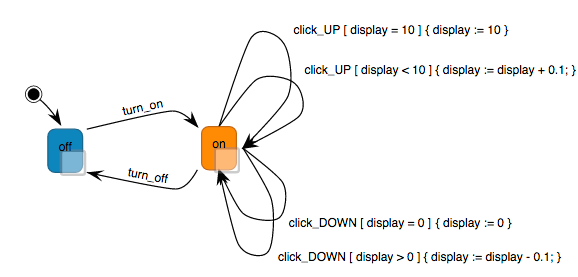}
\caption{\bf Emucharts diagram for the Medtronic MiniMed 530G data entry system.}
\label{fig:medtronic}
\end{figure}

The declarations are followed by the function prototypes of the two
utility functions \emph{enter} and \emph{leave}, the \emph{init} function,
and, for each trigger, one permission and one transition function.
The functions receive a pointer to a structure of type \emph{state} passed
by a calling program.  The \emph{enter} and \emph{leave} functions, called
by the \emph{init} and transition functions, update the \emph{curr\_node}
and \emph{prev\_node} fields, respectively, with the target and source
node label of the executed transition.  The \emph{leave} function has been
introduced to allow future versions to implement checkpointing algorithms.
The \emph{init} function initialises the state's context fields with the
values of the context variables specified in the Emucharts diagram, and the
\emph{curr\_node} field with the label of the initial node.  As mentioned
above, each permission function checks if the current node has a transition
labelled by the respective event.  Then, the matching transition function
chooses among the transitions triggered by that event, according to the
respective guards (assumed to be mutually exclusive).

The implementation file contains the function definitions.  For example,
consider the Emucharts diagram of the data entry system of the Medtronic
MiniMed 530G System shown in Figure~\ref{fig:medtronic}.
The diagram has a context variable \emph{display} of type \emph{double}
represented on 64 bits, which holds the value shown on the device's display.
The node labels and the \emph{state} type are defined as

\begin{small}
\begin{verbatim}
typedef enum { off, on } node_label;
typedef struct {
    D_64 display;
    node_label curr_node;
    node_label prev_node; } state;
\end{verbatim}
\end{small}

The code for the permission function associated with the \emph{click\_UP}
trigger is

\begin{small}
\begin{verbatim}
UC_8 per_click_UP(const state* st) {
    if (st->current_state ==  on) {
        return true;
    }
    return false;
}
\end{verbatim}
\end{small}

\noindent
where the return type UC\_8 (eight-bit unsigned character) is used to represent
the Boolean type.  The transition function is

\begin{small}
\begin{verbatim}
state click_UP(state* st) {
    assert(st->current_state ==  on);
    assert(st->display < 10 || st->display == 10);
    if (st->display < 10 && st->current_state == on) {
        leave(on, st);
        st->display = st->display + 0.1f;
        enter(on, st);
        assert(st->current_state == on);
        return *st;
    }
    if (st->display == 10 && st->current_state == on) {
        leave(on, st);
        st->display = 10.0f;
        enter(on, st);
        assert(st->current_state == on);
        return *st;
    }
    return *st;
}
\end{verbatim}
\end{small}

A proof of the correctness of this translation schema is shown in
Appendix~\ref{corr}.

\section{Case study}
\label{sec:example}

The Alaris~GP, made by Becton Dickinson and Company, was used as a case
study for the MISRA~C code generator.

\begin{figure}[t]
\centering
\includegraphics[width=0.5\linewidth]{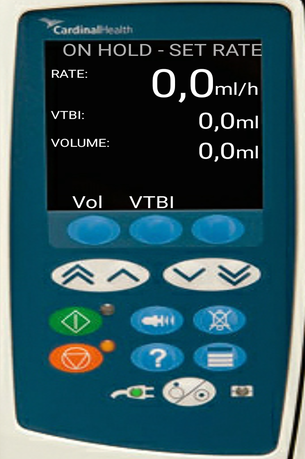}
\caption{Front panel of the Alaris GP infusion pump.}
\label{alarisfp}
\end{figure}

This volumetric infusion pump is a medical device used for controlled
automatic delivery of fluid medication or blood transfusion to patients,
with an infusion rate range between 1~ml/h and 1200~ml/h.  It has
a monochrome dot matrix display with three significant digits, and has 14
buttons for operating the device (see Figure~\ref{alarisfp}).  The pump
has a rather complex user interface, with different modes of operation and
ways of entering data, including the possibility of choosing from a list
of preloaded treatments.  For simplicity, in this paper only the essential
part of the data entry interface, concerning numerical input and display,
is considered.

\begin{sloppypar}
Numerical input is done through the chevrons buttons: upward and downward
chevrons increase and decrease, respectively, the displayed value.  The amount
by which the value is increased or decreased depends on whether a single or
double chevron is pressed, and on the current displayed value.  More
precisely, the displayed value is changed as follows:
\point(i) If the displayed value is below $100$, the value changes by $0.1$
units for a single chevron, and steps up or down to the next decade for
a double chevron (e.g., from $9.1$ to $10.0$);
\point(ii) if the displayed value is between $100$ and $1,000$, the value
changes by $1$ unit for a single chevron, and steps up or down to a value
equal to the next hundred plus the decade of the displayed value for a double
chevron (e.g., from $310$ or $315$ to $410$);
\point(iii) if the displayed value is $1,000$ or above, the value changes by
$10$ units for a single chevron, and steps up or down to a value equal to the
next hundred for a double chevron (e.g., from $1,010$ or $1,080$ to $1,100$).
\end{sloppypar}

\begin{figure}[t]
\centering
\includegraphics[width=1.0\linewidth]{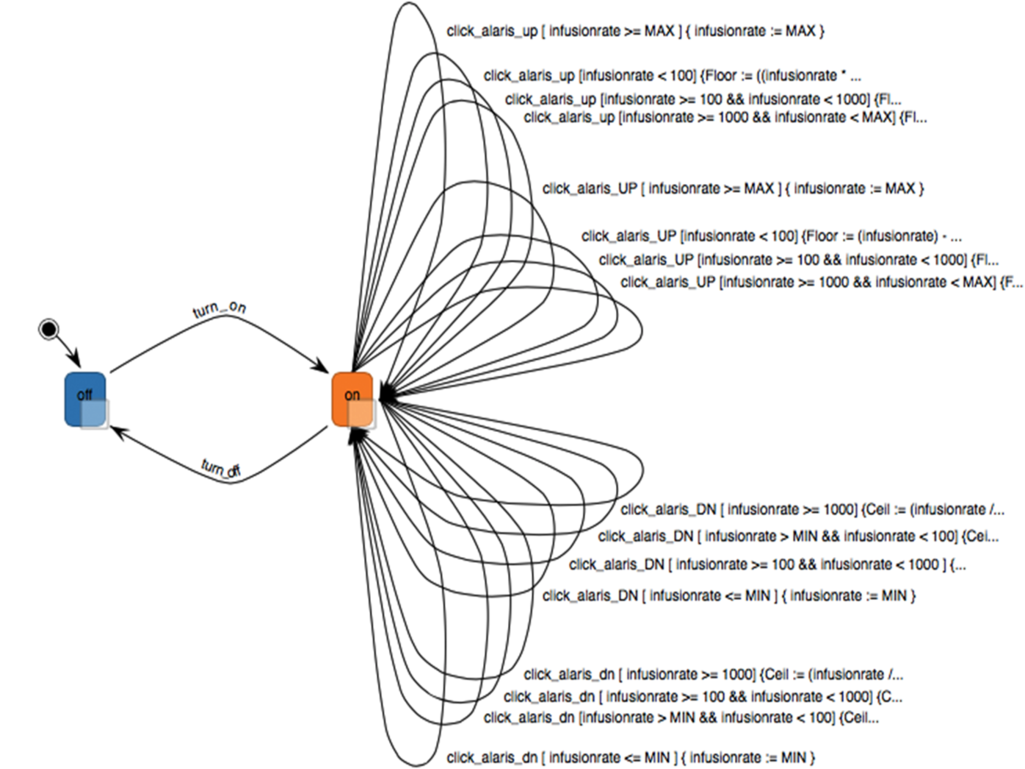}
\caption{Emucharts diagram for numeric data entry.}
\label{alarisemu}
\end{figure}

The Emucharts diagram for the numeric data entry is shown in
Fig.~\ref{alarisemu}.  Triggers \emph{click\_alaris\_up} and
\emph{click\_alaris\_dn} represent clicks on the upward and downward
single-chevron buttons, respectively, and triggers \emph{click\_alaris\_UP} and
\emph{click\_alaris\_DN} represent clicks on the double-chevron ones.  For each
event, combinations of guards and actions specify the rules described above.

The PVS code generator translates the diagram into an executable logic theory,
and the C~code generator produces permission and transition functions for each
trigger, as explained previously.

\subsection{Mobile applications}
\label{aplt}

The PVSio-web framework uses a standard web interface to integrate its tools:
this approach offers a uniform interface that a developer can access with
any web browser.

Our framework has been extended by providing the possibility to run simulations
on a mobile device.  Smartphones and tablets improve usability and help make
user interaction similar to actual device operation.  For example, mobile
devices could be used in a hospital environment to train medical personnel
and patients.

An interactive device can be simulated using the C source code produced by
the PVSio-web generator, compiled and linked with a mobile device-specific
application.  For example, the code for the user interface of the Alaris
infusion pump has been ported to the Android~\cite{chang10} platform using
the Android NDK~\cite{NDK} toolset, which can embed C code in a Java project,
relying on the Java Native Interface (JNI)~\cite{JNI}.

\section{Conclusions}
\label{lausdeo}

We presented the implementation of our MISRA~C code generator for the PVSio-web
prototyping toolkit. Automatic code generation significantly reduces project
development time. Our approach eliminates a human-performed step in the
development process: user interface software engineers no longer
need to convert the design specifications into executable target code. 

Our tool improves the development of safe and dependable user
interfaces, as it greatly facilitates using formal methods easily and reliably
with real UIs, which we demonstrated with the medical device examples in this paper.

Current and future directions include improving this initial integration
with other features of C, still conformant to MISRA~C under the most
recent 2012 rules. We plan to develop code generators for
programming languages such as C++, Java and ADA.

\section*{Acknowledgements}

This work was partially supported by the PRA 2016 project ``Analysis
of Sensory Data: from Traditional Sensors to Social Sensors'' funded by the
University of Pisa.

\Urlmuskip=0mu plus 1mu\relax
\bibliographystyle{eptcs}
\bibliography{mauro}

\appendix

\section{Correctness of code generation}
\label{corr}

In order to assess the correctness of the generated code, the Emucharts
diagram is taken as the reference model, and a correspondence is established
between the evolution of the model and that of the executed code.

\subsection{Transition system for an Emucharts diagram}
\label{emuts}

As discussed above (section~\ref{emupvs}), an Emucharts diagram is a graph of
nodes and labelled transitions, extended with a set of typed context variables,
each one with an initial value.  Its semantics is given by a transition system.
Let the following be defined:

\begin{itemize}\raggedright
\item
A set $N = \{ n_1, \ldots, n_i \}$ of nodes;
\item
a set $X = \{ x_1, \ldots, x_j \}$ of context variables (for simplicity,
assumed to be typeless);
\item
a set $\mathbb{V}$ of values;
\item
a set $E = \{ \epsilon_1, \ldots, \epsilon_k \}$ of events;
\item
a set $G = \{ g_1, \ldots, g_l \}$ of guards, i.e., Boolean expressions
involving variables, constants from $\mathbb{V}$, arithmetic and relational
operators;
\item
a denumerable set $V$ of \emph{valuations}, i.e., functions from $X$ to
$\mathbb{V}$;
\item
a set $A = \{ a_1, \ldots, a_l \}$ of arcs, i.e., 5-tuples of the form $(s, t,
e, g, v)$, where $s, t \in N$ are the arc's source and target node, $e \in E$,
$g \in G$, and $v \in V$ is the valuation defined by the action labelling the
corresponding transition in the diagram; more precisely, $v$ is the valuation
obtained by overriding the previous valuation with the assignments in the
action associated with the arc;
\item
a set $\mathbb{Q}$ of states of the form $\langle n, v \rangle $, with
$n \in N$ and $v \in V$;
\item
a transition relation ${\rightarrow} \subseteq Q \times Q$, defined by
the semantic rules in Figure~\ref{opdiagram}, where the premises contain an
event $\epsilon$, an arc label, and a logical condition, and the consequences
contain a member of the transition relation that is enabled if the condition
holds.
\end{itemize}

With the above definitions, the associated transition system $T$ is the
tuple \mbox{$(\mathbb{Q}, \rightarrow, q_0)$}, where $q_0 = \langle n_0,
v_0 \rangle $ is the initial state.  Since the diagram is deterministic,
given a sequence of event occurrences $e_1,~\ldots, e_k, ~\ldots$, the
transition system has only one sequential path.  If an event cannot affect
a state (either it is not permitted or no guard prefixed by the event is
satisfied), the system does not change state. The operational semantics are
given in Figure~\ref{opdiagram}.

\begin{figure}[t]
\[
\mathbf{arc}~ \frac{\epsilon, (p, q, e, g, v');~ \epsilon = e \land n = p
                    \land v \models g}%
                  {\langle n, v \rangle \rightarrow \langle q, v' \rangle}
\]
\[
\mathbf{idle} ~\frac{\epsilon, (p, q, e, g, v');~ \epsilon \neq e \vee n \neq p
                     \vee v \not\models g}
                   {\langle n, v \rangle \rightarrow \langle n, v \rangle}
\]
\caption{Emucharts operational semantics.}
\label{opdiagram}
\end{figure}

\subsection{Transition system for the generated code}
\label{codets}

The generated functions are used within a more complex system, which is
responsible for catching events at the real or simulated user interface and
for calling the respective functions according to an appropriate protocol:
the \emph{init} function must have been called previously, then, when an event
is catched, the permission function of the corresponding trigger is called,
and only if it returns \emph{true} can the respective transition function
be executed.

Assume that the data entry subsystem of the device is controlled by a program
$P$ that responds to input events by calling the respective functions.
These function will take the device to the next state.


Also the program $P$ can be modelled as a transition system $T_P$ based on the
following sets, each one being isomorphic ($\cong$) to the corresponding set
in $T$, or an extension to that set:
\point(i) A set $N_P \cong N$ of node labels, each represented by an
enumerator of the \emph{node\_label} type in $P$;
\point(ii) a set $X_P = X_c \cup \{x_\mathrm{curr}\}$ of variables,
where $X_c \cong X$, each variable in $X_c$ represents a context field of the
\emph{state} structure in $P$, and $x_\mathrm{curr}$ represents the
\emph{curr\_node} of the \emph{state} structure;
\point(iii) a set $\mathbb{V}_P = \mathbb{V}_c \cup N_P$ of values, where
$\mathbb{V}_c = \mathbb{V}$;
\point(iv) a set $E_P \cong E$ of events, each one associated with one
permission function and one transition function in $P$;
\point(v) a set $G_P \cong G$ of guards, each implemented as the condition of
an \emph{if} statement in $P$;
\point(vi) a denumerable set $V_P = V_c \cup V_n$ of valuations from $X_P$ to
$\mathbb{V}_P$, where $V_c \cong V$ and $V_n \colon
\{ x_\mathrm{curr}\} \rightarrow N_P$;
\point(vii) a set $A_P \cong A$ of arcs, where each arc has the form
$(v_n(x_\mathrm{curr}), v'_n(x_\mathrm{curr}), e, g, v'_c)$, and each arc
represents an \emph{if} statement in the transition function for event
$e$ having guard $g$ as its condition and valuation $v' = v'_n \cup v'_c$
as its controlled statement, with $v'_n$ implemented by the \emph{enter}
function and $v'_c$ by the assignments specified in the Emucharts diagram.

With the above definitions, let $Q_P$ be a set of states where
each state is a pair $\langle v_n, v_c \rangle$, with $v_n \in V_n$, $v_c \in
V_c$.  The transition relation ${\cxsemfreccia{P}} \subseteq Q_P \times Q_P$
is defined by the semantic rules in figure~\ref{oppgm} applied to elements
of the above sets, and implemented by the permission functions, which check
for each event $e$ if the condition $v_n(x_\mathrm{curr}) = p$ holds or not,
and by the transition functions, which check if the current values of the
variables satisfy the guards, and update node and variables accordingly.
The associated transition system $T_P$ is the tuple \mbox{$(Q_P,
\cxsemfreccia{P}, q_{P 0})$}, where $q_{P 0}$ is the state defined by the
initial values of $x_\mathrm{curr}$ and of the context variables, set by
the \emph{init} function.  The operational semantics are given in
Fig.~\ref{oppgm}.

\begin{figure}[t]
\[
\mathbf{arc}_P ~~ \frac{\epsilon, (p, q, e, g, v'_c);~ \epsilon = e
                       \land v_n(x_\mathrm{curr}) = p \land v_c \models g}%
                    {\langle v_n, v_c \rangle ~\cxsemfreccia{P}~
                        \langle q, v' \rangle}
\]
\[
\mathbf{idle}_P ~~ \frac{\epsilon, (p, q, e, g, v'_c);~ \epsilon \neq e
                      \vee v_n(x_\mathrm{curr}) \neq p \vee v_c \not\models g}%
                     {\langle v_n, v_c \rangle ~\cxsemfreccia{P}~
                        \langle n, v_c \rangle}%
\]
\caption{Generated code operational semantics.}
\label{oppgm}
\end{figure}

\subsection{Equivalence of the transition systems}
\label{eqvts}

To prove the correctness of the generated code, we introduce the
definition of equivalence between Emucharts states and the program states.

\noindent
\textbf{Definition 1.}\textit{
A member $m$ of one of the sets $N$, $X$, $\mathbb{V}$, $E$, defined in $T$, is
equivalent ($\sim$) to the member $m_P$ paired to $m$ by the isomorphism
between the set containing $m$ and the corresponding set in $T_P$.}

\noindent
\textbf{Definition 2.}\textit{
A state $q = \langle n, v \rangle$, $q \in Q$,
is equivalent ($\sim$) to a state $q_P = \langle v_n, v_c \rangle$, $q_P \in
Q_P$ iff $n \sim v_n(x_\mathrm{curr})$ --- {so the value of
$x_\mathrm{curr}$ is equivalent to node $n$}, and
$\forall_{x \in X} v(x) = v_c(x_P)$ (i.e., matching variables in
$q$ and $q_P$ have the same values).}

The proof of correctness for the generated code is by induction on the
length of computation.  We assume that $T$ and $T_P$ are the transition
systems modelling, respectively, an Emucharts diagram and a program that uses
the generated code, respecting the previously introduced protocol, and accepts
a sequence of input events.

\noindent
\textbf{Theorem 1.}\textit{
Let $T$ and $T_P$ be the transition systems introduced in the above paragraphs,
and $e = e_1, e_2 \ldots$ be a sequence of input event sequences.  Let $\sigma
= q_0, q_1, \ldots$ and $\sigma_P = q_{P 0}, q_{P 1}, \ldots$ be sequences
of states, with $q_i \rightarrow q_{(i+1)}$ and $q_{P i} \cxsemfreccia{P}
q_{P (i+1)}$.}

We prove that, at each step of the computation, $q_i \sim q_{P i}$:

\noindent
\textbf{Induction base.} $q_0 \sim q_{P 0}$ by construction.\\
\textbf{Induction step.} Let $q_j \sim q_{P j}$ at step $j$.  On the
occurrence of an event $e$, let $q_j \rightarrow q_{(j+1)}$ and $q_{P j}
\cxsemfreccia{P} q_{P (j+1)}$.  We can prove that $q_{(j+1)} \sim q_{P
(j+1)}$ by case analysis: (1) $e$ not permitted in $q_j$; (2) $e$ permitted
and guard not satisfied; and (3) $e$ permitted and guard satisfied.

\noindent
{\bf Case 1: $e$ not permitted}.
If the event is not permitted in the current state, rules~$\mathbf{idle}$ and
$\mathbf{idle}_P$ apply to $T$ and $T_P$, respectively, so that $q_{(j+1)} =
q_j$ and $q_{P (j+1)} = q_{P j}$, equivalent by induction hypothesis.  Recall
that the permission function for $e$ returns \emph{false} in this case, and
by hypothesis program $P$ does not call the corresponding transition function.

\noindent
{\bf Case 2: $e$  permitted and guard not satisfied}.
Also in this case, rules~\textbf{idle} and $\mathbf{idle}_P$ apply to the
transition systems.  The \emph{if\/} statements in $P$ check that the guard
does not hold, and the respective controlled statements are not executed.

\noindent
{\bf Case 3: $e$  permitted and guard satisfied}.
In this case,
Rules~\textbf{arc} and $\mathbf{arc}_P$ apply to both transition systems,
therefore 
\point(i) $T$ moves from state $q_j = \langle n, v \rangle$ to state
$q_{(j+1)} = \langle n', v' \rangle$, or 
\point(ii) $T_P$ moves from state
$q_{P j} = \langle v_n, v_c \rangle$ to state $q_{P (j+1)} = \langle v'_n,
v'_c \rangle$.  Valuation $v'_n$ maps $x_\mathrm{curr}$ to a node label
equivalent by definition to $n'$, and $v'_c$ maps the context variables in
$T_P$ to values equivalent by definition to those assigned by $v'$ to the
context variables in $T'$\@.  

The new states in the two transition systems are therefore equivalent.

\end{document}